# Beyond Credibility: Understanding the Mediators Between Electronic Word-of-Mouth and Purchase Intention

Bogdan Anastasiei
University Alexandrul Ioan Cuza, Iasi, Romania
Faculty of Economics and Business Administration
Dept. of Management, Marketing and Business Administration
**abo28@yahoo.com**

Nicoleta Dospinescu
University Alexandrul Ioan Cuza, Iasi, Romania
Faculty of Economics and Business Administration
Dept. of Management, Marketing and Business Administration
**dnicole@uaic.ro**

Octavian Dospinescu
University Alexandrul Ioan Cuza, Iasi, Romania
Faculty of Economics and Business Administration
Dept. of Accounting, Business Informatics and Statistics
**octavian.dospinescu@feaa.uaic.ro**

## Abstract

This study examines the mediating role of five variables in the relationship between eWOM credibility and purchase intention. Data were collected via an online survey of 359 Romanian participants and analyzed using Structural Equation Modeling. The results show that perceived product quality, emotional response to eWOM and perceived buying risk fully mediate this relationship, while brand trust and brand engagement do not. These findings suggest that effective eWOM communication should enhance quality perceptions, evoke positive emotions, and boost customer confidence by reducing perceived purchasing risks.

## Introduction and Hypotheses Development

In a world in which the customer is always at the center of the business, understanding the mechanisms that determine the attitude towards buying intent is extremely important and has major implications for commerce. Given that an important part of commercial activities are carried out through mechanisms that involve the use of online technologies, the opinions of online customers have a strong impact on purchasing decisions.

Various models have been developed in the literature that consider the direct influence between eWOM and purchase intention, but mediation effects have not been fully covered by previous studies. Based on previous results from the literature, we will determine the factors with mediating potential between eWOM and intention to purchase, thus filling a research gap that will be identified and justified in the following paragraphs. Starting from this research gap, the research goal of our article is to clearly determine whether the relationship between eWOM credibility and purchase intention is directly or mediated by other variables.

Based on the research goal, the main research questions we are considering are the following: is the relationship between eWOM credibility and intention to purchase a direct one or is it mediated by other factors? If the relationship is mediated, what are the mediating factors and what is the nature of this relationship?

In the following, we develop a research model that includes several research hypotheses based on scientific results from the literature. Then, we will analyze the research model using the Structural Equation Modeling technique on a representative sample of respondents. Based on the results obtained from the model analysis, we present the relevant discussions about the hypotheses and the model. The practical implications of our original research results are also presented objectively. The article ends with conclusions, limitations and potential future directions.

**eWOM Credibility**

eWOM refers to recommendations and reviews distributed in the digital space, which have the potential to influence consumer perceptions of products and services. Unlike traditional WOM, eWOM has a much wider audience and is accessible indefinitely (Iqbal, Khan, & Faridi, 2022). In this dynamic, credibility plays a key role, as consumers tend to pay more attention to information that is considered trustworthy and relevant.

The credibility of eWOM is influenced by several factors, including source expertise, perceived honesty, message relevance and information quality. A study conducted by (Mainolfi & Vergura, 2022) emphasizes that consumers' perception of source credibility directly influences the acceptability of the message conveyed. If a review or recommendation is perceived as authentic and based on real experiences, the likelihood of it positively influencing purchase intention increases significantly. According to (Nistor, Moca, Moldovan, Oprean, & Nistor, 2021), building indicators for measuring the sentiment of people on a specific topic of discussion is a very difficult task.

In addition to message characteristics, the platform on which the eWOM is distributed plays an important role in building credibility. Established platforms, such as Amazon, TripAdvisor or major social networks, provide a more secure and transparent environment for consumers, which contributes to increasing trust in the information distributed. However, recent research, including (Reyes-Menendez, Saura, & Martinez-Navalon, 2019), highlights that information overabundance may diminish the eWOM effect, as consumers become more skeptical when encountering conflicting reviews.

An interesting mechanism is related to the formation of positive attitudes towards brands and products. Positive reviews, backed by sources that are considered credible, contribute to creating a favorable image of a brand, which can lead to an increase in purchase intention. For example, a study by (Ngarmwongnoi, Oliveira, AbedRabbo, & Mousavi, 2020) shows that reviews that contain clear details, concrete examples and an authentic tone are more effective in shaping consumer behavior than those that are vague or generic.

Cultural differences also play a significant role in how consumers perceive and use eWOM. In collectivist cultures, where the group has a central role, online reviews are often more influential than in individualist cultures, where consumers tend to be more skeptical.

**Brand Trust**

In today's digital context, where social networks and rating platforms play a key role in purchasing decisions, the concept of eWOM (electronic Word-of-Mouth) has gained significant importance. In the age of digitization, consumer behavior is profoundly influenced by the information available online, especially through electronic Word-of-Mouth (eWOM) communication. The credibility of this information plays a key role in shaping purchase intention, but one factor that mediates this relationship is brand trust. Understanding how brand trust interacts with eWOM trustworthiness to influence purchase intention provides valuable insights for marketing research and brand strategies. Studies, such as the one conducted by (Erkan & Evans, 2016), highlight that a positive perception of eWOM credibility directly influences consumer trust in the product or service being analyzed. However, purchase intention is not only shaped by eWOM credibility, but also by deeper factors, such as the emotional and cognitive connection between the consumer and the brand. Brand trust is defined as consumers' expectation that the brand will deliver on its promises and deliver products or services to its stated standards.

This is based on a trust relationship built over time between the brand and consumers. According to a study by (Jain, Dixit, & Shukla, 2022), brand trust mediates the relationship between the credibility of information received through eWOM and purchase intention by reducing uncertainty and creating a sense of security associated with the purchase decision. The mediation works mainly through how consumers interpret the information received. Where a brand is already perceived as trustworthy, credible eWOM messages become a reinforcing factor. Consumers tend to accept information as authentic and relevant when it is congruent with the existing brand image. For example, the study by (Nurhasanah, Lucky, & Yananto, 2021) demonstrated that for brands with a high degree of trustworthiness, positive reviews generate a greater impact on purchase intention than for brands that are less well known or considered less trustworthy. On the other hand, for brands with weaker reputations, eWOM credibility plays an important role in diminishing skepticism. When the reviews are detailed, well-argued and come from sources perceived as neutral, they can compensate for the initial lack of trust in the brand and stimulate purchase intention.

In this case, eWOM credibility serves as a temporary proxy for brand trust, as concluded by (Seo & Park, A Study on the Influence of the Information Characteristics of Airline Social Media on e-WOM, Brand Equity and Trust, 2018). The relationship between these variables is not without complications. Brand trust can be affected by discrepancies between eWOM messages and direct consumer experience. If messages deemed credible turn out to be misleading, brand trust suffers a sharp decline, which in the long run affects purchase intention. Therefore, in order to ensure a positive relationship between eWOM credibility and buying intention, it is essential for brands to support a consistent experience that validates the positive information distributed online. According to (Seo, Park, & Choi, The Effect of Social Media Usage Characteristics on e-WOM, Trust, and Brand Equity: Focusing on Users of Airline Social Media, 2020), this relationship emphasizes the importance of building a strong brand image and providing products or services that meet consumer expectations.

Taking into consideration the above mentioned aspects, we formulate the following research hypothesis:

***H1: The relationship between eWOM credibility and buying intention is mediated by brand trust.***

**Perceived Product Quality**

Consumer behaviour is influenced to a considerable extent by the opinions expressed online, and eWOM credibility is recognized as a key factor in determining buying intention. However, this influence is not direct and may be mediated by consumers' perception of perceived product quality. Analysis of this complex relationship provides insight into consumer decision-making processes.

Credibility of eWOM is defined as the level of trust given to information and opinions shared online by other users. According to research, including that by (Evgeniy, Lee, & Roh, 2019), consumers tend to place particular importance on reviews that are considered authentic and detailed. This credibility influences the decision-making process by reducing the uncertainty associated with purchasing a product. However, the relationship between eWOM credibility and purchase intention is conditioned by how consumers perceive the quality of the respective products.

Perception of product quality plays an important role in mediating this relationship, as consumers use information from reviews to form an evaluation of the product. Perceived quality, defined as the consumer's subjective evaluation of a product's benefits and performance, is often constructed based on information received through eWOM. For example, a study by (Daowd, et al., 2021) suggests that when reviews are perceived as credible, consumers associate this information with a more favorable perception of product quality, which in turn increases purchase intention.

The mechanism by which quality perception mediates the relationship between eWOM credibility and buying intention involves several steps. First, consumers assess the credibility of eWOMs through the details, source and consistency of information. When these reviews are considered authentic, they positively influence the perception of the product, generating a higher level of confidence in its quality. This increased perception of quality translates into an increased likelihood that the product will be purchased. For example, the study by (Hussain, Ahmed, Jafar, Rabnawaz, & Jianzhou, 2017) emphasizes that reviews that provide specific examples and are supported by real experiences are more effective in shaping quality perception and, hence, purchase intention. Some studies (Furtună, Reveiu, Dârdală, & Smeureanu, 2016) revealed that there are different typologies of consumers and patterns of their behavior, with influence on perception of quality. According to (García-Machado, Figueroa-García, & Jachowicz, 2020), there are many motivational forces that drive consumer choice.

On the other hand, if the information in the eWOM is considered to be unreliable or inconsistent, this can negatively affect the perception of product quality. Thus, consumers become more skeptical and less willing to make the decision to purchase the product. It is important to note that this relationship is also influenced by contextual factors such as product price, brand reputation or product complexity. For example, an expensive or technical product requires a higher level of eWOM credibility to generate a positive perception of its quality, according to research by (Bushara, et al., 2023).

Based on previous studies, it can be argued that perception of product quality plays a key role in mediating the relationship between eWOM credibility and purchase intention. Recent studies demonstrate that the influence of online reviews is not only direct, but also indirect through the way they shape consumers' expectations of the product. Thus, we formulate the following research hypothesis:

*H2: The relationship between eWOM credibility and buying intention is mediated by perceived product quality.*

**Emotional Response to eWOM**

The credibility of reviews, perceived as the authenticity and reliability of the messages conveyed, plays a particularly important role in influencing purchasing decisions. However, the relationship between eWOM credibility and purchase intention is mediated by several factors, including consumers' emotional response to these messages. Recent studies (Liu, Jayawardhena, Osburg, Yoganathan, & Cartwright, 2021) highlight how emotions generated by eWOM content contribute to perceptions and thus to decision-making.

The credibility of eWOM is essential for the messages delivered to have a significant impact on consumers. Information considered authentic and relevant is more likely to trigger positive emotional responses, such as enthusiasm, trust or anticipated satisfaction. These emotional responses play an important media role, transforming consumers' cognitive appraisals into motivational impulses that lead to purchase intent. For example, a study by (Donthu, Kumar, Pandey, Pandey, & Mishra, 2021) shows that positive emotional responses to authentic and well-argued reviews significantly increase the likelihood that a consumer will consider the advertised product.

The emotional response to eWOM can vary depending on the tone of the message and its content. Positive reviews, characterized by enthusiasm and detailed descriptions, tend to trigger favourable emotions, which increase purchase intention. On the other hand, negative reviews, even if perceived as credible, may generate emotions such as frustration, anxiety or disappointment, negatively influencing purchase intention. Research by (Gerrath, Mafael, Ulqinaku, & Biraglia, 2023) emphasizes that the intensity of the emotional response also depends on the degree to which consumers identify with the experiences described in the eWOM, which amplifies or diminishes the impact of the message.

Another important aspect is the context in which consumers interact with eWOM. The digital environment, characterized by accessibility and the sheer volume of information, amplifies emotional responses through factors such as visuals, the narrative structure of the review or the overall tone of the message. For example, reviews that include personal stories and details about user experience are more likely to trigger strong emotions. Studies by (Akdim, 2021) suggest that such reviews not only influence the perception of the product, but also create a deeper emotional connection with the message conveyed.

The mediation of emotional response is also supported by the type of product or service discussed. Products with a strong emotional component, such as those in the luxury or entertainment categories, are more likely to generate intense emotional responses, which increases the influence of eWOM credibility on buying intention (Pangarkar, Patel, & Kumar, 2023). However, for functional or utilitarian products, emotional response plays a more moderate role, being more influenced by rational evaluations.

Given the previous results, it can be argued that the emotional response to eWOM is a key mechanism through which the credibility of reviews influences purchase intention. Therefore, we formulate the following research hypothesis in the model proposed in this paper:

*H3: The relationship between eWOM credibility and buying intention is mediated by emotional response to eWOM.*

**Perceived Purchasing Risk**

The credibility of eWOM is seen as a key determinant of purchase intention, being associated with the level of trust that consumers place in the information offered online. However, the relationship between eWOM credibility and purchase intention is often influenced by intermediary factors, a key one of which is the perceived risk of purchase. Perceived risk plays a mediated role through its influence on how consumers evaluate available information and make purchasing decisions.

The credibility of eWOM is fundamental to consumer trust in online reviews. Information considered authentic and reliable reduces the uncertainty associated with the decision-making process. In contrast, reviews perceived as biased or lacking in detail may amplify the perception of risk, which ultimately discourages purchase intention. Research by (Handoyo, 2024) found that detailed information and consistency of messaging help reduce perceived risk, as consumers view these reviews as valuable sources of information for anticipating product performance.

Perceived purchase risk is defined as the level of uncertainty and potential for loss that a consumer associates with a transaction. This concept can be divided into multiple dimensions such as financial risk, performance risk or social risk. For example, in the case of an expensive or highly technical product, consumers are more likely to perceive a higher financial risk. The credibility of the eWOM plays a key role in mitigating these risks by providing insights from previous users that validate or challenge product expectations. According to (Filieri, Acikgoz, Ndou, & Dwivedi, 2021), consumers perceive credible reviews as a substitute for missing information, which reduces the perceived level of uncertainty. Studies by (Mircea, Stoica, & Ghilic-Micu, 2022), (Hurbean, 2020), (Gînguță, Munteanu, Ștefea, & Noja, 2023), (Gruia, Bibu, Roja, Dănăiață, & Năstase, 2024) reveal that some technologies like blockchain and artificial intelligence can be used to decrease the risk perception, for both individual and institutional consumers.

The relationship between eWOM credibility and purchase intention is thus indirect and mediated by how consumers assess the risks associated with the purchase. For example, when a review is considered credible, it reduces the fear of poor product performance, leading to an increase in purchase intention. Conversely, reviews perceived as less credible may amplify uncertainty, even for products that were previously well rated in the market. The study by (Putra, Sedera, & Fenitra, 2024) showed that perceived risk acts as a psychological barrier between consumers and online purchases, emphasizing the importance of creating highly credible eWOM content to alleviate these concerns.

In addition to its role in reducing perceived risk, eWOM credibility also influences consumers' trust in the brand, which can help to reduce perceived risk even more effectively. For example, according to (Jiang, et al., 2021), reviews that contain positive and detailed descriptions of the user experience can generate a favorable perception of the brand's reputation while reducing concerns about potential losses. This phenomenon is particularly amplified for new or lesser-known brands, which rely on the credibility of eWOM to build consumer trust.

Arguably, perceived purchase risk may be a key mediator in the relationship between eWOM credibility and purchase intention. Reducing perceived risk through credible reviews enhances consumer confidence, creating a favorable framework for purchase decisions.

Based on the arguments presented above, we hypothesize that:

***H4: The relationship between eWOM credibility and buying intention is mediated by perceived purchasing risk.***

**Engagement with the Brand**

The literature (Srivastava & Sivaramakrishnan, 2021) shows that the relationship between eWOM credibility and purchase intention is not direct, but may be mediated by consumer engagement with the brand. This phenomenon involves the active interaction, emotional attachment and level of dedication that consumers develop towards a brand, which significantly influences how they perceive the information they receive.

The credibility of eWOM directly influences the quality and authenticity of consumer interaction with the brand. Messages deemed authentic and relevant stimulate active consumer engagement, generating both emotional and behavioral responses. Recent research by (El-Baz, Elseidi, & El-Maniaway, 2022) highlights that online reviews perceived as credible lead to users being more likely to interact with the brand through digital channels such as social networks, mobile apps or dedicated websites. This increased engagement helps to strengthen consumers' connection with the brand and facilitates the shift from intent to action.

Engagement with the brand serves as a bridge between eWOM credibility and purchase intention, reinforcing the positive perceptions generated by credible reviews. Engaged consumers are more likely to attribute higher value to eWOM messages, as their frequent interaction with the brand amplifies trust and reduces uncertainty. The study by (AlFarraj, et al., 2021) shows that active engagement with a brand not only enhances perceptions of the products or services on offer, but also increases the likelihood that these perceptions positively influence purchase intention.

A key aspect of this dynamic is how engagement with the brand strengthens the emotional relationship between consumers and the products or services being promoted. Credible eWOM messages have a greater capacity to stimulate interest in brands that promote transparency and authenticity, leading to the formation of lasting emotional attachments. For example, in the case of a detailed review of a premium product, consumers may be motivated to explore more about the brand, which increases interaction and purchase intention. Research by (Abuhjeeleh, Al-Shamaileh, Alkilany, & Kanaan, 2023) emphasizes that emotional engagement not only facilitates purchase intention but also contributes to long-term brand loyalty.

Besides the emotional dimension, consumers' behavioral engagement is also an important mediating factor. This includes activities such as sharing reviews on social networks, commenting on posts or attending events organized by the brand. Credibility of eWOM stimulates such behaviors by providing positive experiences that motivate consumers to actively engage with the brand. The studies by (Chen & Zhao, 2021) and (Noble, Noble, & Adjei, 2012) demonstrated that users who actively participate in online discussions related to brands are more likely to exhibit increased purchase intention, as their interactions help build greater trust in the product and reduce perceived risks.

Engagement with the brand works as a catalyst that amplifies the positive effects of credible reviews through greater emotional and behavioral engagement. Thus, we formulate the following research hypothesis within the proposed research model:

***H5: The relationship between eWOM credibility and buying intention is mediated by the intended engagement with the brand in the social networks.***

Based on the above described research hypotheses, we proposed the research model presented in figure 1.

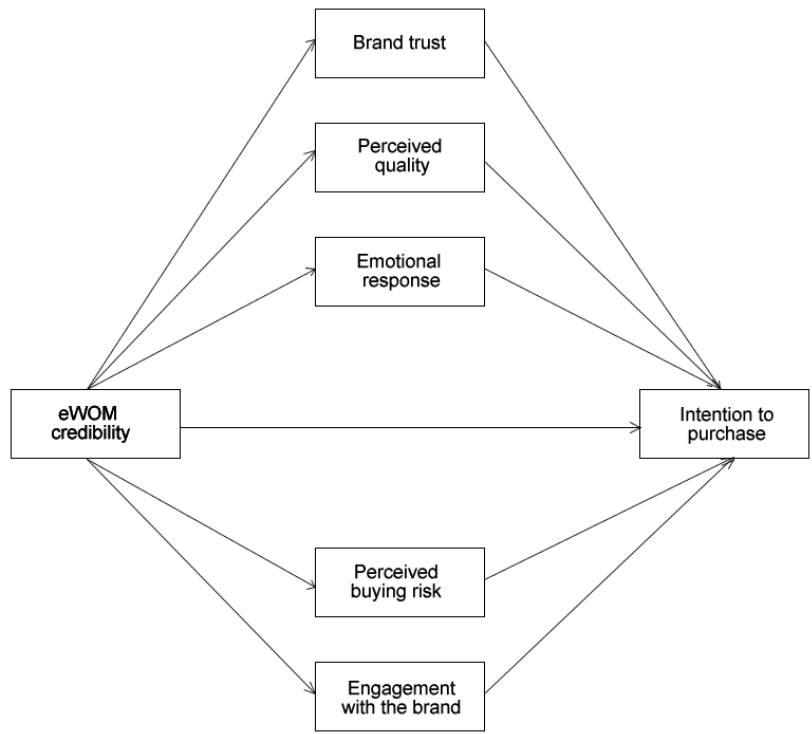

Figure 1 – The research model

## Instrument and methodology

To collect the data for our research we have used an online questionnaire administered to a convenience sample of 359 Romanian students and young professionals. About 58% of the respondents were female, while 42% were male.

The questionnaire was administered online using the Google Forms platform. The questions were translated into Romanian and the answers were obtained in January and February 2025. The respondents were asked to answer forty-four questions, split into seven scales. Each scale measured a specific latent construct: eWOM credibility, intention to purchase, brand trust, perceived product quality, emotional response to message, perceived buying risk and engagement with the brand.

To assess eWOM credibility, the scale of (Ohanian, 1990) was employed. To evaluate the intention to purchase we have used the scale created by (Dodds, Monroe, & Grewal, 1991).

To measure brand trust we have used the scale devised by (Delgado-Ballester, Munuera-Alemán, & Yagüe-Guillén, 2003), while perceived product quality was assessed with the scale developed by (Wu, Lu, Wu, & Fu, 2011). The emotional response was measured using an adapted version of the scale created by (Hecker & Stewart, 1988).

The perceived buying risk was assessed using the scale built by (Kushwaha & Shankar, 2013) and the engagement with the brand was evaluated with the scale devised by (Schivinski, Christodoulides, & Dabrowsk, 2016).

## Data Analysis and Results

Our research procedure was structured in three phases. In the first phase we have executed an exploratory factor analysis (EFA) using the IBM SPSS software, version 26. The goal of the EFA was to establish whether the individual items are appropriately correlated with their constructs. After this analysis, eleven items out of forty-four were removed, because they presented either high cross-loadings or poor loadings. More exactly, we have eliminated one brand trust item, seven emotional response items, one buying risk item and two engagement items. With the remaining items, the Kaiser-Meyer-Olkin indicator for the EFA model was 0.922, showing excellent factor adequacy. The Bartlett's sphericity test was statistically significant (p<0.01).

During the second phase we have run a confirmatory factor analysis (CFA), to evaluate the relationships between our latent variables and their related items. The cutoff values that we have used to estimate the goodness-of-fit of the measurement model were the following: for the $\chi^2$/df ratio – between 1 and 5 (Marsh & Hocevar, 1985), for the comparative fit index (CFI) – 0.900 (Hair, Black, Babin, & Anderson, 2010), for the Tuckey-Lewis index (TLI) – 0.900 (Hu & Bentler, 1999), for the root mean square error of approximation (RMSEA) – 0.08 (Hair, Black, Babin, & Anderson, 2010), for the standardized root mean square residual (SRMR) – 0.08 (Hu & Bentler, 1999), for the goodness-of-fit index (GFI) – 0.800 (Greenspoon & Saklofske, 1998), for the adjusted goodness-of-fit index (AGFI) – 0.800 (Ellison, Steinfield, & Lampe, 2007).

The values for our measurement model are: $\chi^2$/df = 2.192, CFI = 0.927, TLI = 0.919, GFI = 0.843, AGFI = 0.815, RMSEA = 0.058, SRMR = 0.059. All indicators meet the cutoff values, so our model is good fit.

The main indicators of the measurement model are presented in Table 1. All path coefficients are statistically significant (t > 1.96) and their standardized values are higher than 0.5. The average variance extracted (AVE) values are greater than 0.5 for all the latent constructs, indicating good convergent validity. Furthermore, all constructs have good internal consistency (Cronbach's alpha values and composite reliabilities are higher than 0.7).

Table 1. **Summary indicators of the measurement model**

| Constructs and items | Beta | t-value | SE | Alpha | Composite Reliability | AVE |
|---|---|---|---|---|---|---|
| **eWOM credibility** | - | - | - | 0.874 | 0.853 | 0.608 |
| This post is reliable | 0.800 | - | - | - | - | - |
| This post is sincere | 0.774 | 15.904 | 0.068 | - | - | - |
| This post is a safe source of information | 0.718 | 14.482 | 0.070 | - | - | - |
| This post is objective | 0.685 | 13.670 | 0.070 | - | - | - |
| This post is authoritative | 0.546 | 10.469 | 0.078 | - | - | - |
| This post is trustworthy | 0.769 | 15.789 | 0.065 | - | - | - |
| This post is useful | 0.680 | 13.541 | 0.065 | - | - | - |
| **Brand trust** | - | - | - | 0.830 | 0.800 | 0.634 |
| This brand would make any effort to satisfy me | 0.748 | - | - | - | - | - |
| I could rely on this brand to solve my problems | 0.884 | 16.208 | 0.067 | - | - | - |
| This brand would be interested in my satisfaction | 0.705 | 13.056 | 0.069 | - | - | - |
| This brand would compensate me in some way for the problems I might have | 0.631 | 11.598 | 0.078 | - | - | - |
| **Perceived product quality** | - | - | - | 0.864 | 0.879 | 0.754 |
| This product is of good quality | 0.761 | - | - | - | - | - |
| This product is of high workmanship | 0.766 | 14.446 | 0.071 | - | - | - |
| This product is durable | 0.790 | 14.937 | 0.078 | - | - | - |
| This product is reliable | 0.821 | 15.541 | 0.077 | - | - | - |
| This product is dependable | 0.649 | 12.074 | 0.092 | - | - | - |
| **Emotional response** | - | - | - | 0.911 | 0.888 | 0.656 |
| Joyful | 0.792 | - | - | - | - | - |
| Surprised | 0.852 | 18.318 | 0.060 | - | - | - |
| Delighted | 0.895 | 19.588 | 0.060 | - | - | - |
| Enthused | 0.812 | 17.172 | 0.059 | - | - | - |
| Impressed | 0.776 | 16.185 | 0.062 | - | - | - |
| Captivated | | 0.696 | 14.123 | 0.062 | - | - | - |
| Grateful | | 0.593 | 11.661 | 0.068 | - | - | - |
| **Perceived buying risk** | | - | - | - | 0.840 | 0.793 | 0.576 |
| I am afraid that this product might not work properly | | 0.858 | - | - | - | - | - |
| I am afraid that I could lose money if I buy this product | | 0.847 | 17.212 | 0.057 | - | - | - |
| I am afraid that this product might be harmful | | 0.686 | 13.742 | 0.059 | - | - | - |
| I am afraid that this product will not fit in my self-image or self-concept | | 0.624 | 12.216 | 0.065 | - | - | - |
| **Engagement with the brand** | | - | - | - | 0.937 | 0.937 | 0.808 |
| I would comment on the posts of this brand | | 0.946 | - | - | - | - | - |
| I would comment on the videos of this brand | | 0.971 | 42.468 | 0.023 | - | - | - |
| I would comment on the photos of this brand | | 0.946 | 37.989 | 0.025 | - | - | - |
| I would share the posts of this brand | | 0.720 | 18.204 | 0.047 | - | - | - |
| **Intention to purchase** | | - | - | - | 0.878 | 0.859 | 0.723 |
| The likelihood of buying this product is (very low – very high) | | 0.915 | - | - | - | - | - |
| The likelihood of using this product is (very low – very high) | | 0.784 | 18.691 | 0.047 | - | - | - |
| My desire to have this product is (very low – very high) | | 0.826 | 20.391 | 0.048 | - | - | - |

At the end of the second phase, we had to assess the discriminant validity of our measurement model. To that effect, we have compared the construct squared correlations with the average variance extracted. As it results from Table 2, all AVE values (on the main diagonal) are greater than the squared correlations, which suggests good discriminant validity.

Table 2. **Average variance extracted and squared correlations**

| | eWOM credibility | Brand trust | Prodct quality | Emotional response | Perceived buying risk | Engagement | Intention to purchase |
|---|---|---|---|---|---|---|---|
| eWOM credibility | **0.608** | | | | | | |
| Brand trust | 0.616 | **0.634** | | | | | |
| Product quality | 0.236 | 0.243 | **0.754** | | | | |
| Emotional response | 0.465 | 0.425 | 0.259 | **0.656** | | | |
| Perceived buying risk | 0.026 | 0.036 | 0.024 | 0.017 | **0.576** | | |
| Engagement | 0.232 | 0.144 | 0.002 | 0.163 | 0.000 | **0.808** | |
| Intention to | 0.441 | 0.411 | 0.279 | 0.479 | 0.108 | 0.145 | **0.723** |

| | | | | | | |
|---|---|---|---|---|---|---|
| purchase | | | | | | |

During the third phase of our analysis we investigated our causal model, presented in Figure 1. The goodness-of-fit indicators for this model were as follows: $\chi^2/df = 2.273$, CFI = 0.920, TLI = 0.914, GFI = 0.831, AGFI = 0.806, RMSEA = 0.060, SRMR = 0.069. These values indicate good model fit. The direct effects in the causal model can be examined in Table 3 and Figure 2.

Table 3. **Summary of direct effects**

| Path | Coefficient | p |
|---|---|---|
| eWOM credibility -> Intention to purchase | 0.188 | 0.159 |
| eWOM credibility -> Brand trust | 0.843 | < 0.01 |
| Brand trust -> Intention to purchase | 0.152 | 0.119 |
| eWOM credibility -> Perceived quality | 0.415 | < 0.01 |
| Perceived quality -> Intention to purchase | 0.273 | < 0.01 |
| eWOM credibility -> Emotional response | 0.806 | < 0.01 |
| Emotional response -> Intention to purchase | 0.365 | < 0.01 |
| eWOM credibility -> Perceived buying risk | -0.215 | < 0.01 |
| Perceived buying risk -> Intention to purchase | -0.201 | < 0.01 |
| eWOM credibility -> Engagement with the brand | 0.565 | < 0.01 |
| Engagement with the brand -> Intention to purchase | 0.101 | < 0.05 |

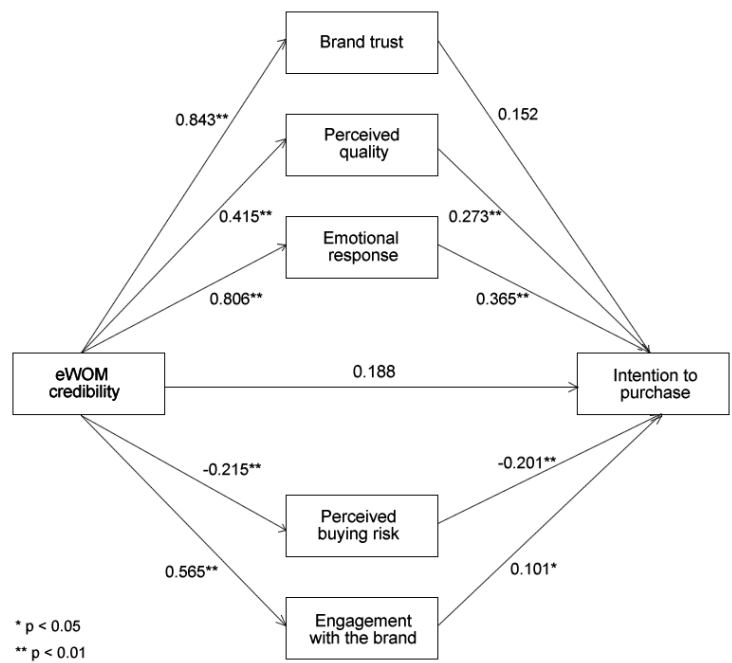

Figure 2. **Direct effects – path coefficients**

The mediation analysis summary can be found in Table 4.

Table 4. **Mediation analysis summary**

| Hypothesys | Relationship | Direct effect | Indirect effect | p | Conclusion |
|---|---|---|---|---|---|
| H1 | EC → BT → IP | 0.188 (p = 0.159) | 0.128 | 0.089 | No mediation |
| H2 | EC → PQ → IP | | 0.113 | 0.001 | Full mediation |
| H3 | EC → ER → IP | | 0.294 | < 0.001 | Full mediation |
| H4 | EC → PR → IP | | 0.043 | 0.010 | Full mediation |
| H5 | EC → EB → IP | | 0.057 | 0.050 | No mediation |

EC – eWOM credibility, IP – intention to purchase, BT – brand trust, PQ – product quality, ER – emotional response, PR – perceived buying risk, EB – engagement with the brand

The results revealed significant indirect effects for the following variables: perceived product quality, emotional response to eWOM and perceived buying risk. In conclusion, these variables have a significant mediating role between eWOM credibility and intention to purchase. Furthermore, the direct effect of credibility on intention to purchase was found nonsignificant (p = 0.159). Hence, the above-mentioned variables fully mediate the relationship between eWOM credibility and purchase intention.

These findings will be closely examined in the following section.

## Discussions

Our findings indicate that the relationship between eWOM credibility and purchase intent is not direct but mediated by other variables. One of these variables is perceived product quality. This suggests that credible eWOM influences the likelihood of buying a product by shaping the perception of its quality, rather than having a direct effect. Previous studies carried out by (Yu, Kangmun, & Taewoo, 2019) and (Roy, Datta, Mukherjee, & Basu, 2021) support our conclusion, showing that, on the one hand, positive eWOM improves perceived product quality, whereas, on the other hand, higher perceived quality is associated with stronger buying intent. All these findings align with the signaling theory in marketing, highlighting the role of eWOM messages as infomational cues that help customer to assess product quality even in the absence of direct experience.

Another mediator between eWOM credibility and buying intent is the emotional response to the eWOM message. As a consequence, eWOM triggers the desire to purchase the product only by eliciting a positive emotional reaction like joy, enchantment, surprise, arousal or enthusiasm. Put differently, high eWOM credibility translates into stronger emotional engagement that creates the urge to buy. This result is consistent with the findings of (Hecker & Stewart, 1988), who demonstrated the role of emotions as mediators between advertising content and consumer response to advertising. More recently, (Ruiz-Mafe, Bigné-Alcañiz, & Currás-Pérez, 2020) showed that online reviews that provoke positive emotions like pleasure or arousal are more likely to increase the readers' intentions to follow the reviewer's advice and purchase the product.

Furthermore, our research shows that perceived buying risk fully mediates the liaison between eWOM credibility and intention to purchase. Credible eWOM reduces the perceived risks associated with the purchase (especially social and financial risks), thus increasing the probability of buying the product. When customers find a reliable and authoritative eWOM source, they feel more confident in their purchase decision, minimizing uncertainty, hesitation and fear. These findings are in line with the results obtained by authors like (Amarullah, Handriana, & Maharudin, 2022), who pointed out the role of perceived risk in the relationship between eWOM and buying intention.

As evidenced by our research, brand trust and engagement with the brand on social media are not mediators between eWOM credibility and buying intention. While eWOM may enhance customers' trust in the brand, this trust alone does not convert into a higher propensity to purchase the product. This finding seems to contradict some prior research that shows that trust is a key driver of buying intent (Zhao, Wang, Tang, & Zhang, 2020). While our study does not discard the importance of brand trust, it reveals that customers' purchase decisions are influenced more by their direct evaluation of product quality and purchasing risks rather than by overall brand confidence. These factors have a stronger effect on purchase intent because they address fundamental concerns like product reliability or potential loss from buying, which carry more weight for the customers than the trust they place in the brand.

Moreover, though high eWOM credibility can lead to increased social media interaction with the brand, this interaction does not necessarily influence the customers' purchasing decisions. Customers may consider eWOM information a valuable decision-making support tool, without feeling the need to interact further with the brand through liking, sharing or commenting on online posts. While engagement can foster brand awareness or loyalty, it does not immediately translate into stronger purchase intentions. Social media engagement could just reflect sheer interest or curiosity rather than willingness to buy.

In other words, this finding highlights that online engagement does not necessarily play a pivotal role in converting eWOM credibility into higher likelihood of buying. The influence of eWOM credibility is conveyed by the perceived product quality and perceived purchasing risks, rather than being amplified by consumer engagement with the brand.

## Limitations and further research

This study presents several limitations. In the first place, the convenience sampling method was employed, which might reduce the generalizability of our results. Furthermore, our sample is made up of Romanian students only, with the majority being aged under 30. We have chosen students because they actively participate in social networks and regularly interact with their peers.

Since this research assessed purchase intention rather than actual behavior, future investigations could use field studies to explore whether our mediating variables similarly mediate real-world purchasing decisions. Additionally, further research might consider potential moderators in the relationships between mediators and purchase intention. These moderators could be customer's familiarity with the product, customer's personality traits, product category or product price. That could offer greater insight into the underlying mechanisms through which different variables affect purchase decisions.

# Conclusions and practical implications

eWOM trustworthiness has a strong impact on customers, determining their purchase intentions (Leong, Loi, & Woon, 2022). The primary contribution of this study lies in demonstrating that the influence of credibility on buying intent is not direct, but rather mediated by other key factors: perceived product quality, perceived buying risk and emotional reaction of the eWOM recipient. Consequently, a well-crafted eWOM message should address issues of both product quality and buying risks. By directly tackling these critical factors, such a message can reduce customer doubt, build trust and improve the likelihood of influencing purchasing behavior. eWOM communication should not only highlight the product's attributes and benefits but also provide reassurances that mitigate potential risks, thereby fostering a favorable customer response.

From a practical standpoint, companies should prioritize strategies that encourage the production of credible eWOM messages that create strong quality perceptions about the product. This includes leveraging verified customer reviews, expert endorsements, and crafting emotionally compelling messages that resonate with receivers. In addition, businesses should actively focus on people concerns about purchasing risks by providing transparent information, guarantees and customer testimonials, ensuring a well-rounded approach that strengthens the impact of eWOM on customer decisions.

**Appendix**

The survey body text:
*Suppose you are browsing Facebook and come across the following post written by a virtual friend, regarding a laptop they recently purchased.*
*"I promised myself I wouldn't be the type of person who praises gadgets online, but this laptop is truly worth talking about. It's the kind of device that makes your life easier and more organized, and that's no small thing.*
*Let me start with what I liked the most: its performance. It's incredibly fast and handles multitasking perfectly. I can work, follow a webinar, and have several tabs open simultaneously without any issues. The battery is also a pleasant surprise – it lasts me through an entire workday. Plus, the design is simple and elegant.*
*Another aspect I really liked is the screen. The colors are vibrant, and the details are sharp, making any activity much more enjoyable. Additionally, the keyboard is so comfortable that it completely transformed my typing experience. The hours spent working are no longer a chore.*
*The only thing that slightly bothered me is that the speakers aren't exactly great. If you want to listen to music or watch a movie, the sound quality is decent but not spectacular. However, considering how well it performs in all other areas, this minor inconvenience doesn't bother me too much.*
*In the end, I can say that it's more than just a laptop – it's a reliable partner. Whether you're working, studying, or simply looking for a moment of relaxation, it does its job very well. I would make the same choice anytime!"*
*Please read the text above carefully and then answer the following questions as accurately as possible.*

The questions of the survey:
*The post above is…*
(5-point agreement-disagreement scale for each statement)
… trustworthy

… a reliable source of information
… one you can rely on when making a purchase decision
… one that comes from an authorized source
… objective, impartial
… in good faith
… useful, helpful

*Based on the post, indicate to what extent you agree with each of the following statements.*
(5-point agreement-disagreement scale for each statement)
*The manufacturer of this laptop…*
… would be sincere and honest in addressing my concerns.
… would make every effort to ensure my satisfaction.
… is a company I could count on.
… would be interested in my satisfaction.
… would compensate me if I had a problem with the product.

*Based on the post, the laptop being discussed seems to be…*
(5-point agreement-disagreement scale for each attribute)
… of good quality
… very well-made
… durable
… reliable
… solid, robust

*Please specify your agreement or disagreement with the following statements.*
(5-point agreement-disagreement scale for each statement)
*When I read this post, I felt…*
… happy, joyful
… pleasantly surprised
… delighted
… irritated
… excited
… impressed
… confused
… skeptical
… captivated
… amused
… calm, relaxed
… grateful
… indifferent
… informed, clarified

*Please tell us what you would do if you encountered content created by the laptop brand mentioned in the post on a social media platform.*
(5-point agreement-disagreement scale for each statement)

I would "like" the brand's posts.
I would "like" the brand's videos.
I would write comments on the brand's posts.
I would write comments on the brand's videos.
I would write comments on the brand's images or photos.

I would share the brand's posts.

*Indicate to what extent you agree with the following statements.*
(5-point agreement-disagreement scale for each statement)
*If I were to buy the laptop mentioned in the post…*
… I am afraid it might not work well
… I am afraid I would regret the money spent
… I am afraid it might be dangerous or unsafe
… I am afraid it might not suit me or my personality
… I am afraid other people would disapprove or laugh at me

*Please indicate whether you would buy and use the laptop discussed in the post.*
*(5-point bipolar scale, very low – very high, for each question)*
The likelihood of purchasing this laptop is…
The likelihood of ever using this laptop is…
My desire to own this laptop is…